\documentclass[reprint,aps]{revtex4-2}

\usepackage{graphicx}

\usepackage[utf8]{inputenc}
\usepackage[T1]{fontenc}
\usepackage{upgreek}

\usepackage{amsmath, amsthm, amssymb}

\begin{document}
	\title{A Mott-Schottky Analysis of Mesoporous Silicon in Aqueous Electrolyte by Electrochemical Impedance Spectroscopy}
	
	%
	%
	%
	%
	%
	\author{Manuel Brinker}
	\email{manuel.brinker@tuhh.de}
	\affiliation{Hamburg University of Technology, Institute for Materials and X-Ray Physics, 21073 Hamburg, Germany}
	\affiliation{Center for X-Ray and Nano Science CXNS, Deutsches Elektronen-Synchrotron DESY, 22607 Hamburg, Germany} 
	\author{Patrick Huber}
	\email{patrick.huber@tuhh.de}
	%
	%
	\affiliation{Hamburg University of Technology, Institute for Materials and X-Ray Physics, 21073 Hamburg, Germany}
	\affiliation{Center for X-Ray and Nano Science CXNS, Deutsches Elektronen-Synchrotron DESY, 22607 Hamburg, Germany} 
	%
	%
	%
	\begin{abstract}
		\noindent Nanoporosity in silicon leads to completely new functionalities of this mainstream semiconductor. In recent years, it has been shown that filling the pores with aqueous electrolytes in addition opens a particularly wide field for modifying and achieving active control of these functionalities, e.g., for electrochemo-mechanical actuation and tunable photonics, or for the design of on-chip supercapacitors. However, a mechanistic understanding of these new features has been hampered by the lack of a detailed characterization of the electrochemical behavior of mesoporous silicon in aqueous electrolytes. Here, the capacitive, potential-controlled charging of the electrical double layer in a mesoporous silicon electrode (pore diameter $7\,\mathrm{nm}$) imbibed with perchloric acid solution is studied by electrochemical impedance spectroscopy. Thorough measurements with detailed explanations of the observed phenomena lead to a comprehensive understanding of the capacitive properties of porous silicon. An analysis based on the Mott-Schottky equation allows general conclusions to be drawn about the state of the band structure within the pore walls. Essential parameters such as the flat band potential, the doping density and the width of the space charge region can be determined. A comparison with bulk silicon shows that the flat band potential in particular is significantly altered by the introduction of nanopores, as it shifts from $1.4\pm0.1\,\mathrm{V}$ to $1.9\pm0.2\,\mathrm{V}$. Overall, this study provides a unique insight into the electrochemical processes, especially the electrical double layer charging, of nanoporous semiconductor electrodes.  
	\end{abstract}
	\maketitle
	\section{Introduction}
	\noindent Porous silicon is a highly versatile material with a plethora of possible application fields. Porous silicon exhibits outstanding characteristics concerning fundamental properties and combines these in research for applications in a wide variety of research fields. These include medical and biological sciences, where porous silicon has been studied with respect to sensor applications~\cite{Lin1997,Vendamani2022,Sailor1997}, optics and electronics~\cite{Wu2008,Micera2022} and a tightly adjustable control of a drug's release kinetics~\cite{Tzur2015}. In fundamental research, porous silicon represents an excellent host material to study the confinement of matter inside the pores, with respect to its structure and dynamics.\cite{Gruener2008,Calus2012, Gor2015,Kondrashova2017,Huber2015,Vincent2016,Cencha2020}\\
	Furthermore, a mechanical actuation in porous silicon induced by humidity or gas sorption has been investigated.\cite{Zhao2014,Ganser2016,Fratzl2016,Dolino1996} As a contributing effect of liquid-adsorption measurements, a deformation of porous silicon is revealing porous silicon's mechanics.\cite{Grosman2015,Gor2015,Gor2017,Rolley2017} A highly reversible deformation, or more precisely straining, of porous silicon in a controlled manner can be achieved either by an electrosorption-induced change in surface stress of the inner surface area of porous silicon~\cite{Brinker2021} or by the incorporation of the electro-active polymer polypyrrole into the pore space of porous silicon~\cite{Brinker2020,Brinker2022}. These methods have the benefit of operating under electrochemical control and thus the reversible straining is highly adjustable by the applied potential.\\
	Another field of research that employs electrochemical properties of porous silicon is the broad investigation of porous silicon as an application as a lithium battery anode material~\cite{Jiang2014a,Jia2020,Cheng2023}.\\
	Porous silicon can be synthesized in an electrochemical etching procedure in hydrofluoric acid. The details of the synthesis have been studied in great detail.\cite{Canham2015a,Lehmann2002,Sailor2011,Zhang2007,Chen2003} In particular electrochemical impedance spectroscopy (EIS) has been employed to gain a deeper understanding of the synthesis process in hydrofluoric acid solutions~\cite{Ronga1991,Searson1991,Popkirov1997,Parkhutik2000b,Husairi2014,Mogoda2019}. Investigations in aqueous and especially acidic environments focus mostly on an anodic oxidation of porous silicon~\cite{Mula2014,Parkhutik2000,Bsiesy1991}. Therefore, the focus here is set on an EIS investigation of the electrochemical properties of mesoporous silicon, with a pore diameter of $7\,\mathrm{nm}$ (hereafter referred to as simply porous silicon), in aqueous acidic solution. Rather than exploring an electrochemical redox reaction~\cite{Mei2018b}, in this publication the reversible charging of the electric double layer (EDL) in a porous silicon electrode is in the focus.\\
	The evaluation of the porous silicon's capacitive features is conducted in terms of the Mott-Schottky equation~\cite{Sato1998}
	\begin{equation}
		\label{eq_Mott-Schottky}
		c_\mathrm{SCR}^{-2}=-\frac{2}{eN_\mathrm{a}\varepsilon_\mathrm{r}\varepsilon_0} \left(E-E_\mathrm{fb}-\frac{k_\mathrm{B}T}{e}\right),
	\end{equation}
	where $c_\mathrm{SCR}$ denotes the capacitance of the space charge region (SCR) on the silicon side of the interface, $N_\mathrm{a}$ the semi conductor's doping concentration, $\varepsilon_\mathrm{r}$ and $\varepsilon_0$ the relative and vacuum permittivity, respectively. By $e$, $k_\mathrm{B}$ and $T$ is referred to the elementary charge, the Boltzmann constant and the temperature. The Mott-Schottky equation \ref{eq_Mott-Schottky} conveniently connects the physical property of the SCR capacitance on the one side of the semi-conductor-electrolyte interface to the electrochemical property of applied potential $E$ on the other, solution side of the interface. In a simple experiment, $c_\mathrm{SCR}$ is measured in dependence of the applied, static potential $E$ so that in a plot of $c_\mathrm{SCR}^{-2}$ versus $E$ the slope of the linear relation of the two
	\begin{equation}
		\label{eq_slope}
		\mathrm{d}c_\mathrm{SCR}^{-2}/\mathrm{d}E=-2/eN_\mathrm{a}\varepsilon_\mathrm{r}\varepsilon_0
	\end{equation}
	yields the doping concentration $N_\mathrm{a}$.\cite{Gelderman2007} Equation \ref{eq_slope} therefore requires the capacitance normalized to the electrode surface. Furthermore, the sign of the slope indicates the doping type, either n- or p-type.\cite{Zhang2007} Next, the flatband potential $E_\mathrm{fb}$ indicates the transition from the depletion regime, where the SCR develops, to the accumulation regime, where the SCR is annihilated by an opposite band bending at the interface. The flatband potential $E_\mathrm{fb}$ can be determined in the plot of $c_\mathrm{SCR}^{-2}$ versus $E$ by the crossing point with zero, i.e.\ at $c_\mathrm{SCR}^{-2}=0$, where $E=E_\mathrm{fb}+k_\mathrm{B}T/e$. The width $w$ of the SCR is obtained by~\cite{Grundmann2010}
	\begin{equation}
		\label{eq_scr-w}
		w= \sqrt{-\frac{2\varepsilon_\mathrm{r}\varepsilon_0}{e N_\mathrm{a}} (E-V_\mathrm{fb}-\frac{k_\mathrm{B}T}{e})}. 
	\end{equation}
	\section{Experimental}
	Porous silicon is synthesized in an electrochemical anodization procedure. A constant current is applied between a bulk silicon wafer and a platinum counter electrode in a hydrofluoric acid electrolyte in a PTFE etching cell. The starting material is lightly doped p-type silicon (Si-Mat Silicon Materials GmbH) it has a thickness of $100\pm10\,\mathrm{\mu m}$, a (100) orientation and a resistivity of $0.01\,-\,0.02\,\mathrm{\Omega cm}$, which corresponds to a doping level of approximately $3\cdot10^{18}$ to $8\cdot10^{18}\,\mathrm{cm^{-3}}$~\cite{Lehmann2002}. The current applied accounts to a density of $12.5\,\mathrm{mA cm^{-2}}$. The electrolyte is a 2:3 volumetric mixture of hydrofluoric acid a (48\%, Merck Emsure) and ethanol (absolute, Merck Emsure). A scanning electron micrograph of the porous silicon layer yields a thickness of $630\,\mathrm{nm}$. A nitrogen sorption isotherm measurement provides an average pore diameter of $3.36\,\mathrm{nm}$ and a porosity of $54\,\mathrm{\%}$ for the porous silicon layer. Electrochemically stable porous silicon is needed to perform an EIS analysis. In an acidic aqueous electrolyte solution the porous silicon's inner surface area is oxidized by an applied potential. Therefore, a constant potential of $1.2\,\mathrm{V}$ is applied for 20 hours. Thus, the oxidative currents are suppressed and solely capacitive currents remain.\cite{Brinker2021} All in all, the porous silicon electrode has an electrochemically active inner surface area of $164.7\pm0.4\,\mathrm{cm^2}$.\\
	EIS is performed in a three electrode setup, utilizing a commercial reversible hydrogen reference electrode (RHE, Gaskatel HydroFlex), in an electrolyte solution consisting of $1\,\mathrm{mol\,l^-1}$ perchloric acid (70\% $\mathrm{HClO_4}$, Merck Suprapur, diluted with deionized water). EIS employs an alternating, sinusoidal potential $E$. It consists of a static potential $E_\mathrm{s}$ that stays constant for the duration of the measurement and a sinusoidal potential. The amplitude $E_0$ of the sinusoidal potential is comparatively small. Here $10\,\mathrm{mV}$ are used. The resulting harmonic current response $J_\mathrm{s}$ is measured between WE and CE. The potentiostat (Metrohm PGSTAT204)is equipped with an impedance analysis module (Metrohm Autolab FRA32) to perform EIS measurements. The frequency is swept over several decades to perform an EIS analysis of the electrochemical processes occurring at the porous silicon interface. The impedance is recorded and averaged at 12 frequency $f$ steps per decade from the kHz regime down to single Hz. The result is displayed in a Nyquist plot that shows $-Z_\mathrm{im}$ versus $Z_\mathrm{re}$. 
	\section{Results \& Discussion}
\begin{figure}[t]
		\centering
		\includegraphics{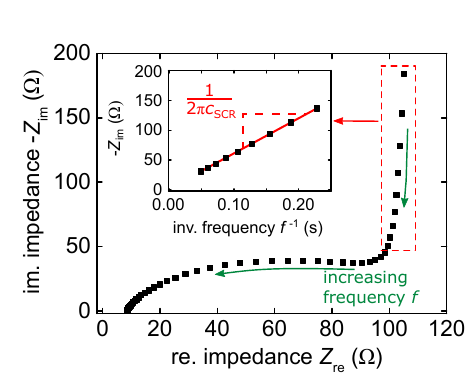}
		\caption{Characteristic Nyquist plot of a porous silicon electrode in $1\,\mathrm{ml^{-1}}$ perchloric acid at a static applied potential of $0.2\,\mathrm{V}$ with frequencies in the range of $100\,\mathrm{kHz}$ to $4.3849\,\mathrm{Hz}$. In the inset the data points of the vertical increase regime at the lowest frequencies (red indication) are plotted versus their inverse frequency $f^{-1}$. A linear fit to the data points yields a capacitance of $c_\mathrm{SCR}=0.21\,\mathrm{mF}$. }
		\label{fig_Nyquist}
	\end{figure}

\begin{figure}[t]
		\centering
		\includegraphics{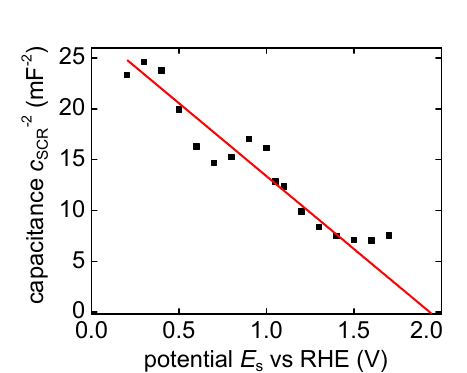}
		\caption{Mott-Schottky plot of a porous silicon electrode. The capacitance determined by EIS is plotted as $c_\mathrm{SCR}^{-2}$ versus $E_\mathrm{s}$, the respective static potential. A linear data fit yields a flatband potential of $V_\mathrm{fb}=1.9\pm0.2\,\mathrm{V}$ and a doping concentration of $N_\mathrm{a}=(3.1\pm0.1)\cdot10^{19}\,\mathrm{cm^{-3}}$.}
		\label{fig_Mott-Schottky}
	\end{figure}

\begin{figure}[t]
		\centering
		\includegraphics{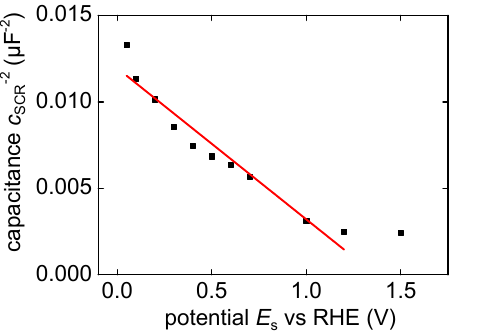}
		\caption{Mott-Schottky plot of a bulk silicon electrode in $1\,\mathrm{ml^{-1}}$ perchloric acid. The capacitance determined by EIS is plotted as $c_\mathrm{SCR}^{-2}$ versus the respective static potential $E_\mathrm{s}$. $c_\mathrm{SCR}$ is determined via EIS measurements with frequencies in the range of $1\,\mathrm{MHz}$ to $500\,\mathrm{mHz}$ and static potentials in the range of $0.05-1.5\,\mathrm{V}$. A linear fit to the Mott-Schottky plot yields a flatband potential of $V_\mathrm{fb}=1.4\pm0.1\,\mathrm{V}$ and a doping concentration of $N_\mathrm{a}=(2.6\pm0.3)\cdot10^{19}\,\mathrm{cm^{-3}}$.}
		\label{fig_Mott-Schottky_bulkSi}
	\end{figure}

	\noindent An exemplary Nyquist plot of a porous silicon electrode in perchloric acid at a static potential of $0.2\,\mathrm{V}$ with frequencies in the range of $100\,\mathrm{kHz}$ to $4.3849\,\mathrm{Hz}$ is depicted in Figure \ref{fig_Nyquist}.\\
	The course of the Nyquist plot closely resembles that of a typical EDL charging process in an electrode. 
	At the highest frequency points, at low values for $Z_\mathrm{re}$, the onset of a half-circle  can be observed. Subsequently, following an intermediate region, at lower $f$ and larger $Z_\mathrm{re}$ the data points approach towards an vertical increase. These features are typical for double layer charging.\cite{Mei2018} They will be further analyzed in the following.\\
	The intersection at $-Z_\mathrm{im}=0$ marks the resistance of the porous silicon electrode~\cite{Mei2018} and it amounts here to $8.38\,\Omega$. Although the half-circle is not particularly pronounced as compared to other electrodes~\cite{Mei2018}, it can be clearly identified. The diameter of the half-circle determines the bulk electrolytes resistance.\cite{Mei2018} It is here $82.1\,\mathrm{\Omega}$, which is in good agreement to literature.\cite{Brickwedde1949} In the vertical regime following the half-circle, the data points converge to the cut-off frequency at approximately $f_\mathrm{c}=4\,\mathrm{Hz}$. The intermediate regime is not distinct, which is an indication that the EDL charging is not inhibited by the resistance of the bulk electrolyte.\cite{Mei2018}\\ 
	In the subsequent vertical regime, solely at the lowest frequencies near $f_\mathrm{c}$, the EDL can be considered as fully charged. An approximation by an equivalent circuit of the Nyquist plot's total course constitutes a difficult endeavor as it is often equivocal and prone to errors introduced by intuition.\cite{Mei2018} However, in the vertical regime a simple RC-circuit can be utilized to approximate the EDL. The resistance can be identified with $Z_\mathrm{re}$, while the capacitance is connected to $c_\mathrm{SCR}$. Thus, $Z_\mathrm{im}$ simplifies to
	\begin{equation}
		\label{eq_C-EIS}
		Z_\mathrm{im}=-\frac{1}{2\pi c_\mathrm{EIS}f},\:\:\:\mathrm{for}\:\:\: f<f_\mathrm{c}.
	\end{equation}
	The data points of the vertical regime near $f_\mathrm{c}$ are further analyzed in a second plot, shown in the inset in Figure \ref{fig_Nyquist}. Here, $-Z_\mathrm{im}$ of these points is plotted versus $f^{-1}$. Their linear dependence is clearly noticeable. The slope of a linear fit to the data is given by $1/{2\pi c_\mathrm{SCR}}$. Therefore, it is possible to determine the EDL capacitance as $c_\mathrm{SCR}=0.21\,\mathrm{mF}$. This is in good agreement to a value of $0.26\,\mathrm{mF}$ determined by cyclic voltammetry.\\
	The described evaluation procedure for $c_\mathrm{SCR}$ is repeated for Nyquist plots with different static potentials in the range of $E_\mathrm{s}=0.2\,\mathrm{V}\,\text{--}\,1.7\,\mathrm{V}$. Thereby, the data for a Mott-Schottky plot is obtained and the result is displayed in Figure \ref{fig_Mott-Schottky}.
	The values for $c_\mathrm{SCR}^{-2}$ are decreasing with increasing potential $E_\mathrm{s}$. The data points decrease in a linear fashion until an ensuing plateau is reached at $1.5\,\mathrm{V}$ and $7\,\mathrm{mF^{-2}}$. However, around $1.0\,\mathrm{V}$ the course deviates from the linear dependence and a feature that could be described as a peak is present. Surface states occurring on the oxidised surface~\cite{Chazalviel1982} of the porous silicon electrode would result in such a deviation to higher $c_\mathrm{SCR}^{-2}$ values~\cite{Zhang2007}. A linear fit to the data in the range up until $1.5,\mathrm{V}$ yields further insights. Firstly, the intersection of the fit with $c_\mathrm{SCR}^{-2}=0$ at the abscissa marks the flatband potential. A value of $V_\mathrm{fb}=1.9\pm0.2\,\mathrm{V}$ is determined. That means, up until $V_\mathrm{fb}$ the porous silicon electrode is in the depletion regime. This value is larger in comparison to $0.38\,\mathrm{V}$, which marks the flatband potential determined in hydrofluoric acid as an electrolyte solution and for, not porous, but bulk silicon with a resistivity four order of magnitude lower.\cite{Searson1991} Porous silicon with a higher conductivity investigated in a non-hydrofluoric aqueous electrolyte might have a higher flatband potential, as seen here.\cite{Ottow1998}\\
	To gain further insight, an equal EIS analysis is performed on bulk silicon. The bulk silicon sample has the same properties as the starting material for the porous silicon synthesis, in particular the same resistivity as stated above. The bulk silicon sample is prepared equally, i.e. the same anodic oxidation step is undertaken. The resulting Nyquist-plots are analyzed in the same fashion as described above so that $c_\mathrm{SCR}$ is determined for different values of $E_\mathrm{s}$ and a Mott-Schottky plot is obtained, depicted in Figure \ref{fig_Mott-Schottky_bulkSi}.
	It has a similar course compared to the Mott-Schottky plot of porous silicon. $c_\mathrm{SCR}^{-2}$ decreases linearly until a plateau starting at $1.2\,\mathrm{V}$ is reached. A linear fit to the data yields a flatband potential of $V_\mathrm{fb}=1.4\pm0.1\,\mathrm{V}$. It is slightly lower in comparison to $V_\mathrm{fb}$ of porous silicon. Thus, porous silicon seems to have an increased flatband potential compared to porous silicon, which has been reported before in literature.\cite{Merazga2019,Ronga1991}\\
	For both the bulk and the porous silicon electrodes a negative slope of the linear fit can be ascertained. Thus, a p-type semiconductor characteristic can be confirmed for the electrodes. Furthermore, the fit's slope yields the doping concentration, see equation \ref{eq_slope}. A doping density of $N_\mathrm{a}=(3.1\pm0.1)\cdot10^{19}\,\mathrm{cm^{-3}}$ is obtained for the porous silicon electrode. The doping density is slightly larger compared to the one stated by the manufacturer. However, the fit of the bulk silicon's Mott-Schottky plot yields a doping density of $N_\mathrm{a}=(2.6\pm0.3)\cdot10^{19}\,\mathrm{cm^{-3}}$, which is in good agreement with the porous silicon electrode.\\
	Finally, the knowledge of the flatband potential and $V_\mathrm{fb}$ and the doping density $N_\mathrm{a}$ enables a computation of the SCR width within the pore walls of the porous silicon electrode. The boundaries of the applied potential $E$ $0.0\,\mathrm{V}$ and $1.5\,\mathrm{V}$ lead to width values of $w=8.9\pm0.2\,\mathrm{nm}$ and $w=4.0\pm0.1\,\mathrm{nm}$. These values seem reasonable as the pore wall thickness is well within this range.
	\section{Conclusion}
	\noindent In summary, we successfully investigated a porous silicon electrode with respect to a capacitive charging of its electric double layer in an aqueous, non-hydroluoric electrolyte solution by technique of electrochemical impedance spectroscopy. The Mott-Schottky analysis yields insight into the energetic band structure within the porous silicon pore walls as it was possible to determine the flatband potential and the width of the SCR of the porous silicon electrode. In particular, the flatband potential is significantly altered as it shifts from $1.4\pm0.1\,\mathrm{V}$ to $1.9\pm0.2\,\mathrm{V}$ by the introduction of pores into a silicon electrode.\\
	Overall, this study provides unique insight into the interaction between the semiconductor properties of porous silicon and its electrochemical features. Here, it could be interesting to further investigate porous silicon's potential in terms of a supercapacitor application~\cite{Westover2014}. Furthermore, research into a hybrid material with a deposited metal layer onto the pore walls of porous silicon could benefit from the presented study. The goal to create such a material combination could be to influence the capacitive or conductive features of porous silicon. This hybrid sample type would establish a semiconductor-metal contact at the interface of the pore wall which, in case of a Schottky contact, resembles the electrolyte-semiconductor interface.
	\clearpage
	\section*{Acknowledgements}  
	\noindent This work was supported by the Deutsche Forschungsgemeinschaft (DFG) within the Collaborative Research Initiative CRC 986 "Tailor-Made Multi-Scale Materials Systems" Project number 192346071. P.H. also acknowledges support by the CRC 1615 "SMART Reactors for Future Process Engineering" Project number 503850735. This work has also received funding from the European Innovation Council (EIC) under the European Union's Horizon 2020 research and innovation program under grant agreement number 964524 EHAWEDRY: "Energy harvesting via wetting-drying cycles with nanoporous electrodes" (H2020-FETOPEN-1-2021-2025). We also acknowledge the scientific exchange and support of the Centre for Molecular Water Science CMWS, Hamburg (Germany).\\
	\textbf{Data and materials availability} \par
	\noindent All data that is needed to evaluate the conclusions is presented in the paper. The raw data of the electrochemical impedance spectroscopy experiments is available at TORE (https://tore.tuhh.de/), the Open Research Repository of Hamburg University of Technology, at the doi: .\\
	\textbf{Author contributions} \par 
	\noindent M.B. and P.H. conceived the experiments. M.B. performed the material synthesis and the electrochemical impedance spectroscopy measurements. M.B. and P.H. performed the data analysis. M.B. and P.H. wrote the manuscript and proofread it.\\ 
	The authors declare that they have no competing interests.\\
	
	\clearpage
	
	%
	
	
\end{document}